# Ferroelectric domain inversion and its stability in lithium niobate thin film on insulator with different thicknesses

Guang-hao Shao,[1] Yu-hang Bai,[1] Guo-xin Cui,[1] Chen Li,[1] Xiang-biao Qiu,[1] De-qiang Geng,[2] Di Wu,[1] and Yan-qing Lu[1,*]

[1]National Laboratory of Solid State Microstructures, College of Engineering and Applied Sciences, and Collaborative Innovation Center of Advanced Microstructures, Nanjing University, Nanjing 210093, China
[2]Jinan Jingzheng Electronics Co., Ltd., Jinan 250100, China

**ABSTRACT:**

Ferroelectric domain inversion and its effect on the stability of lithium niobate thin films on insulator (LNOI) are experimentally characterized. Two sets of specimens with different thicknesses varying from submicron to microns are selected. For micron thick samples (~28 μm), domain structures are achieved by pulsed electric field poling with electrodes patterned via photolithography. No domain structure deterioration has been observed for a month as inspected using polarizing optical microscopy and etching. As for submicron (540 nm) films, large-area domain inversion is realized by scanning a biased conductive tip in a piezoelectric force microscope. A graphic processing method is taken to evaluate the domain retention. A domain life time of 25.0 h is obtained and possible mechanisms are discussed. Our study gives a direct reference for domain structure-related applications of LNOI, including guiding wave nonlinear frequency conversion, nonlinear wavefront tailoring, electro-optic modulation, and piezoelectric devices.

Email: *yqlu@nju.edu.cn

Lithium niobate (LN) has been widely used in many optoelectronic devices for decades. It is even considered to be "the silicon of photonics" because of its versatile electro-optic, nonlinear, pyroelectric, piezoelectric, and acousto-optic properties.[1-3] Recent advances in ferroelectric domain engineering on LN further boosted the research on LN worldwide.[4, 5] Quasi-phase-matching (QPM) frequency conversion, polarization rotation, and macro-phonon polariton excitation have been first predicted and then demonstrated in periodically poled lithium niobate (PPLN).[3, 6-9] The coupling of different physical processes in a domain-engineered LN even gives rise to some unique features such as polarization independent frequency conversion.[10, 11] However, most of the work on LN is still based on bulk samples of 0.5–1 mm in thickness. Owing to the trend of photonic integration, PPLN guiding wave devices have become increasingly important. Although Ti diffusion and proton exchange techniques have been used for years, the severe polarization dependency, extra propagation loss, and large mode size have significantly hampered the applications. Recently, a new platform called lithium niobate thin film on insulator (LNOI) has been proposed and commercially developed, which may open a new window for future ultra-compact active integrated photonic devices and circuits.[12]

LNOI is a thin ion-sliced LN film bonded on a $SiO_2$/LN substrate, which is similar to its silicon on insulator (SOI) counterpart. A thin electrode layer of Au or Pt could be inserted over or under the $SiO_2$ layer. The top LN film could be either x- or z-cut. Because of the unique thin film configuration, the LN film shows interesting advantages in photonic crystal slabs, electro-optic modulators, and microdisk resonators, *etc*.[13-16] Although it is still in its infancy, domain inversion could be an essential research topic for LNOI as well, similar to what happened with bulk LN crystals. So far, very little work has been done on LN thin film poling. Though Gainutdinov *et al.* have initiated the study on domain formation and polarization reversal of LNOI with submicron thickness, only nanometer domains were fabricated and investigated, which are too small for most optical applications.[17] Moreover, in their work, the time interval between poling and testing is only 30 s, which is quite a short time period. The lifetime of domain-inverted regions was not studied, either. Therefore, we believe that domain inversion of LNOI should be investigated with both wider dimension and longer time duration for practical applications. It is known that a submicron guiding wave layer is required to ensure single-mode transmission in high-index LN thin films. In addition, the coherence length for nonlinear frequency conversion is typically in the range of several microns to tens of microns. Therefore, domain inversion of LNOIs with thicknesses up to tens of microns should be more extensively studied in a much longer time period. The domain structures created should also be expanded to at least the micron sale.

In this Letter, domain inversion and domain stability are studied in two types of LNOI samples: micron-scale (~28 μm) and submicron-scale (540 nm) thick LN thin films. Both samples were of +z-cut with the ferroelectric polarization pointing to the sample surface. For the thick sample, metallic electrodes were deposited and patterned on the film surface through lift-off photolithography. Domain reversal was achieved by applying DC high voltage pulses. Domain profile was inspected using a polarizing optic microscope. Conventional $HF:H_2O_2$ etching was also carried out to double-check the domain morphology. For the thinner sample, domains were written by scanning a biased conductive tip over a 4.5-$μm^2$ area in a piezoelectric force microscopy (PFM). Domain stability is examined by measuring the electromechanical response to a small amplitude AC signal on the tip. The lifetime of the reversed domain is then estimated based on the domain area obtained from the PFM phase image.

The schematic diagram of our experiment setup for a few micron-thick LNOI is shown in Fig. 1 (a). The sample consists of four layers (from top to bottom), +z-cut LN thin film (28 μm), Au (100 nm), $SiO_2$ (2 μm), and LN substrate (500 μm), as-provided by Nanoln Electronics (Jinan, China). Since the sample is relatively thick, normal room-temperature poling technique is applicable. Cr comb-shaped electrodes, 100 nm in thickness, were sputter-deposited and patterned via lift-off photolithography. Since the coercive field of LN is 21 KV/mm, ~580 V is the theoretical value for our sample. Ten 620 V positive poling pulses were applied to the Cr electrodes to ensure complete domain penetration as the Au electrode is grounded through silver paste. Thus, the periodically poled LN thin film on insulator (PPLNOI) could be fabricated.

This PPLNOI was first inspected in a reflective polarizing optical microscope. When placed between perpendicular polarizers, straight domains are clearly observed, as shown in Fig. 2(a). Then, the PPLNOI sample was etched in $HF:H_2O_2$ to highlight the domain walls due to differentiate etching. Straight domain walls appear as expected in the optic microscope image shown in Fig. 2(b). The measured domain period is ~80 μm, as defined by the electrode pattern.

In order to check the stability of the inverted domains in such micron-thick LNOI samples, another PPLNOI sample with the same pattern is prepared. The same microscopy and etching characterizations are performed 35 days after poling. As shown respectively in Fig. 2(c) and (d), domains and domain walls can still be observed clearly. The uneven domains in Fig. 2 (d) may be ascribed to the expansion of the negative domains beyond the electrode under a high poling voltage. The results shown in Fig. 2 indicate that the reversed domains in micron-thick LNOIs are as stable as those in LN bulk crystals.

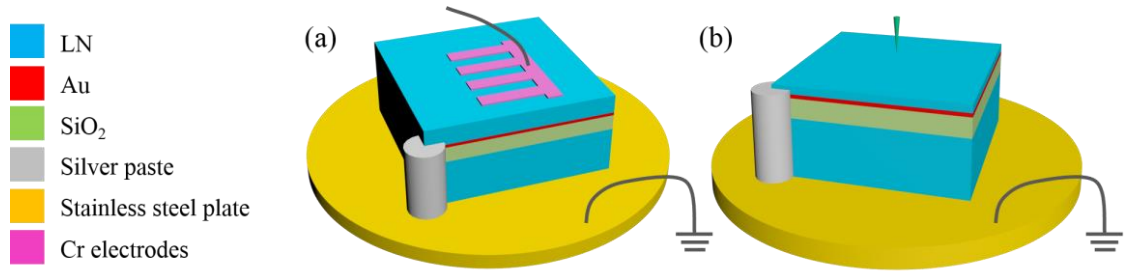

Fig. 1. Schematic diagrams of LNOIs and their corresponding poling techniques for 28-μm (a) and 540-nm (b) thick samples. Blue, red, light green, gray, orange, and pink stand for LN, Pt/Au, SiO$_2$, silver paste, stainless steel plate, and Cr electrodes, respectively. The stainless steel plate is connected to the ground. Positive DC poling voltages are applied to the Cr electrodes in (a) or a conductive tip in (b).

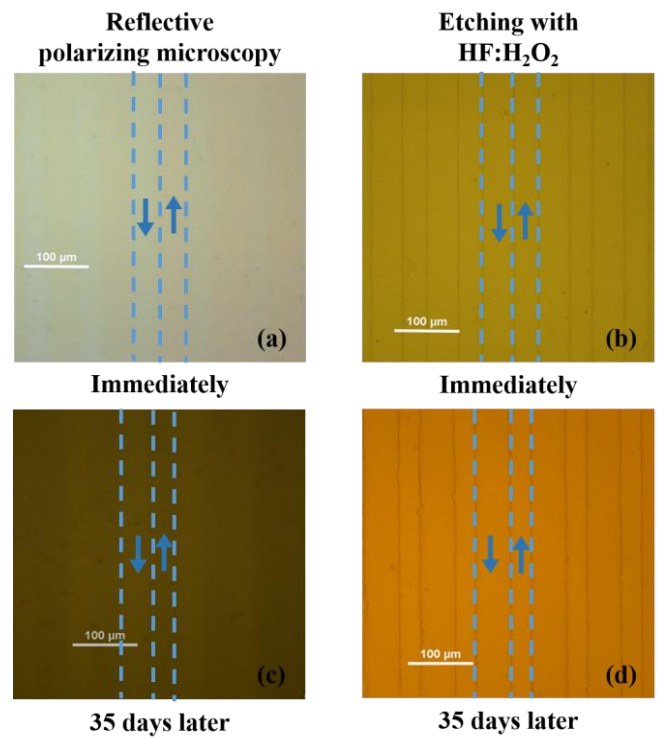

Fig. 2. Domain poling results for 28-μm-thick, 80-μm period PPLNOI. Taken immediately under reflective polarizing microscopy (a), etching with HF:H$_2$O$_2$ (b). Taken 35 days later under reflective polarizing microscopy (c), etching with HF:H$_2$O$_2$ (d), respectively. Up and down arrows indicate positive and negative domains, respectively. Dash lines stand for domain walls.

Though 28-μm samples are not very commonly used for waveguide devices, it is still quite valuable in many other applications, including piezoelectric application, electro-optic modulation, and nonlinear wave front shaping. For example, if a Gaussian beam is injected to a fork-pattern poled LNOI sample, second-harmonic optical vortices could be generated and diffracted back.[18] More applications for complex wave front tailoring could be expected by using properly designed domain patterns.

To the best of our knowledge, 28 μm is the thinnest commercially available micron-scale LNOI so far. Although a 10-μm-thick free-standing LN thin film has been fabricated based on a similar process, it is quite difficult to obtain 10-μm-thick LNOI. According to our sample supplier, the fabricated processes are different for LNOIs with micron and submicron thicknesses. For some unknown reason, the metal electrode layer and the sliced LN thin film itself are quite fragile during micron thick LNOI fabrication if the thickness is less than 28 μm. However, LNOI with a submicron thickness (50 nm to 900 nm) can be achieved with a different fabrication pathway. Therefore, the study of domain inversion and domain stability in LNOI still have a thickness gap from 1 μm to 28 μm. We expect that advances in the LNOI fabrication technology would help to fill the gap soon.

As for sub-micron thick LNOI, we tried but in vain to pole the thin LNOI using patterned electrodes due to the difficulty in focus during photolithography. Domain inversion and domain stability are studied by PFM, instead.[19] Fig. 1(b) shows the schematic diagram of the thinner LNOI sample, provided by the same supplier. The thickness of the +z-cut LN thin film, Pt, $SiO_2$ and LN substrate are 540 nm, 100 nm, 1.7 μm and 500 μm, respectively. The positive poling voltage is applied to the conductive tip scanning over an desired pattern on the sample surface. The optimized poling voltage is 30 V. A 3 μm × 1.5 μm rectangular negative domain is patterned in 4 min.

Domain stability is investigated by measuring the pizeoresponses to 1.0 V AC stimulus applied on the tip at certain spans of time after poling.[20-23] In positive domain areas, the positive voltage compresses the domain, while the negative voltage stretches it. The piezoelectric deformation is opposite in negative domain areas because the piezoelectric coefficient changes sign. Therefore, a 180° PFM phase contrast should be observed in two antiparallel domains. Indeed, in Fig. 3(a), the 180° phase contrast is observed as expected, as measured immediately after poling. Since the original LN thin film is +z cut (in yellow), the poled area (in purple) must be negative, which is similar to micron-thick LNOI and bulk LN. However, the negative domain is not stable, as evidenced in the PFM phase image measured 25 h after poling. As shown in Fig. 3(b), the patterned negative domain is partially switched back, leaving separate nanometer-sized negative domains in the positive matrix.

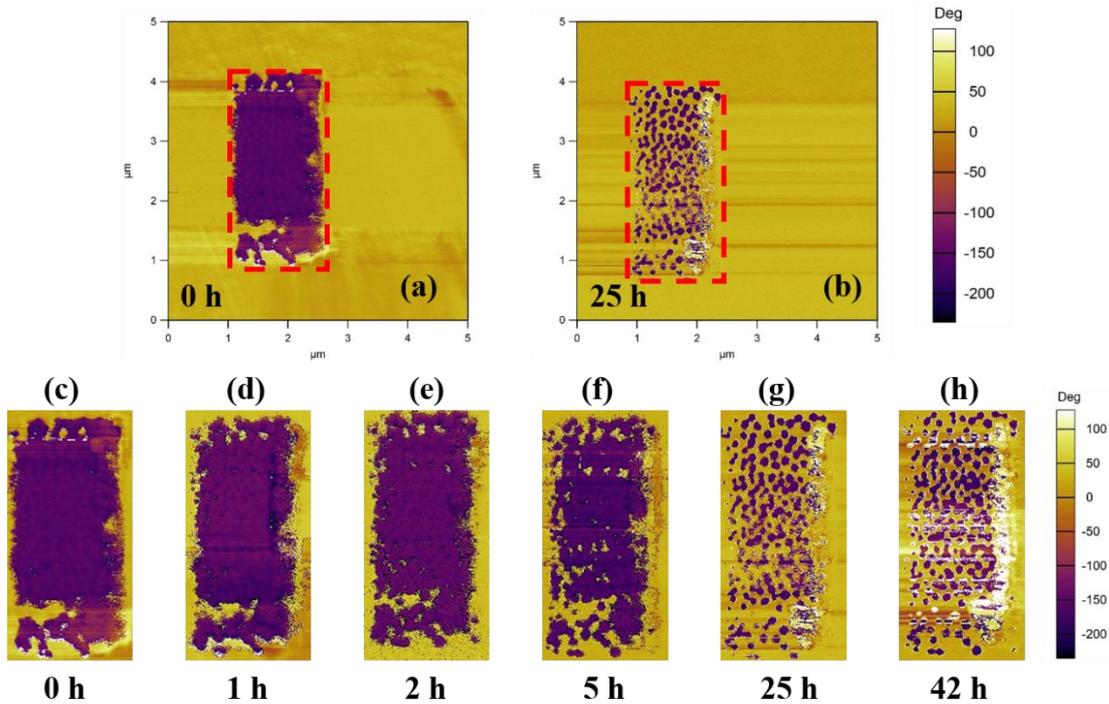

Fig. 3. PFM phase images of the tip-patterned domain. The time intervals between conductive tip scanning and reading are (a) 0 h (immediately) and (b) 25 h. Red rectangles indicate the area over which the positively biased tip scanned. Snapshots of negative domain evolution taken at (c) 0 h, (d) 1 h, (e) 2 h, (f) 5 h, (g) 25 h, and (h) 42 h after poling, respectively.

Gainutdinov *et al.*[17] studied the creation and stability of nanodomains in LNOI. However, the domain size is too small for practical applications. In most applications, larger domains with long retention are desired. The negative domain is checked from time to time to study the domain structure evolution. As shown in Fig. 3(c)-(h), back-switching of the negative domain first appears adjacent to the domain boundary. Then, small positive domains appear in the middle of the negative domain after 2 h. With increasing time, these small positive domains grow and get connected, leaving the rectangular negative domain break into isolated nano-sized islands in the matrix of large and continuous positive domain.

Since these nano-sized negative domains remain even after 42 h, the domains might have penetrated the top LN film. Otherwise, the domain may decay in a short time.[17] However, the negative domain is obviously not very stable, as compared with the thick LNOI. To quantitatively evaluate the domain evolution, the negative domain area in each snapshots shown in Fig. 3(c)-(h) is estimated by counting the purple pixels. Figure 4 shows the negative domain area as a function of retention time. It is observed that ~14.4% of the negative domain written into LNOI switches back to positive polarization within 5 h. Only ~54.6% of the negative domain remains after 25 h. We fit the data with a reduced exponential equation and obtain the relation $P \approx 194378 \times e^{-t/12.475} + 192766$. The residual negative domain area saturates to about ~50% of its original value. If we define a critical domain lifetime at which $1/e^2$ of the original negative domain remains, the obtained lifetime is ~25.0 h. Moreover, our sample is always in the chamber during the measurement under constant environmental conditions. As domain stability is very sensitive to the environment, it may deteriorate as temperature or humidity changes unfavorably.

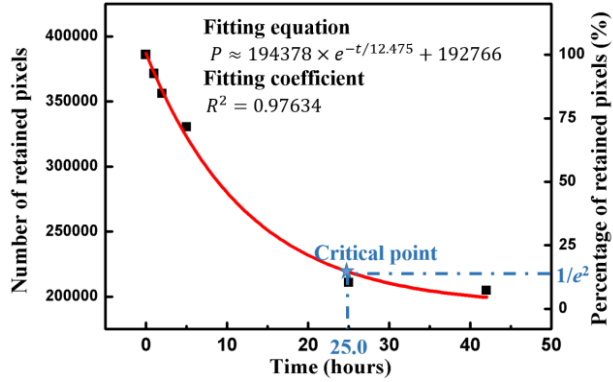

Fig. 4. Retained negative domain area, represented by the number of pixels in the PFM phase images, as a function of time after poling.

In this work, we successfully obtained stable PPLNOIs in a sample with a LN layer 28 μm in thickness. For a thinner LNOI sample, 540 nm in thickness, the lifetime of negative domains is estimated at about 25.0 h. This, however, is much longer than the 0.5 min as reported by Gainutdinov et al.[17] The obtained lifetime might be useful for short-time research and applications employing unique advantages in LNOI. Because of the complexity in LNOI fabrication, many factors may affect the domain stability at different sample thickness. Earlier works have shown that the weak stability in thin LN samples may result from domain electrostatic repulsion.[24, 25] The electrostatic field inside the LN thin film may drive the negative domain to switch back to its original positive state. Besides, according to the sample supplier, the bonding methods in the fabrication processes are different for submicron- and micron-thick LNOI, which may also cause different stability behaviors.

Although retention issues of negative domains have been observed in submicron-thick LNOI, this problem is still solvable as achieved in free standing PPLN thin films.[26] In addition to the fabrication processes, some postprocessing techniques can also be used to stabilize ferroelectric domains in 10-μm or thinner LNOIs. In contrast to the conventional proton exchange or Ti diffusion processes in bulk LN, LNOI may support both TE and TM modes without any extra propagation loss. Therefore, more multifunctional integrated photonic circuits could be expected based on a single LNOI chip containing a nonlinear light source, high speed electro-optic modulators, and other units. Even long-range surface plasmon polariton could be experimentally realized, as the structure of PPLNOI is similar to that designed by Wu, et al..[27] Besides, an optical frequency comb generator could be expected in a poled LNOI microring resonator through the QPM quadratic frequency conversion.[13, 28]

If needed, the conductive tip could supply high enough voltages to pole micron-scale LNOI or even bulk LN. Though taking longer time, the tip-poling process is precise positioning, providing a possible way to selectively pole a particularly small area with designed pattern. Thus it could be an effective method to fabricate artificial defect induced structures within as-fabricated domain patterns for intriguing applications.[29, 30]

In conclusion, two different large-area domain poling techniques were demonstrated for micron- and submicron-thick LNOI samples. Micron-thick PPLNOI was successfully fabricated by electric field application and examined by polarizing optical microscopy and etching. No domain back-switching has been observed in patterned negative domains after 35 days. For submicron-thick LNOI, negative domains can be written by scanning a biased conductive tip over the sample surface. However, the negative domain gradually switches back to the positive domain after poling. The domain lifetime is only 25.0 h. Eventually, about 50% of the negative domain area could be retained. Our study may give some direct guidance for domain-related LNOI applications such as guiding wave QPM nonlinear frequency conversion, nonlinear wavefront tailoring, electro-optic modulation, and piezoelectric devices.

**Acknowledgments**

This work was sponsored by National 973 program under Contracts No. 2012CB921803 and the National Natural Science Foundation of China (Grants No. 61490714 and No. 61225026), and by the Program for Changjiang Scholars and Innovative Research Team at the University under Contract IRT13021. Guang-hao Shao thanks Mr. De-Qiang Geng from Nanoln Electronics (Jinan, China) for his fruitful discussion.

**Figures**

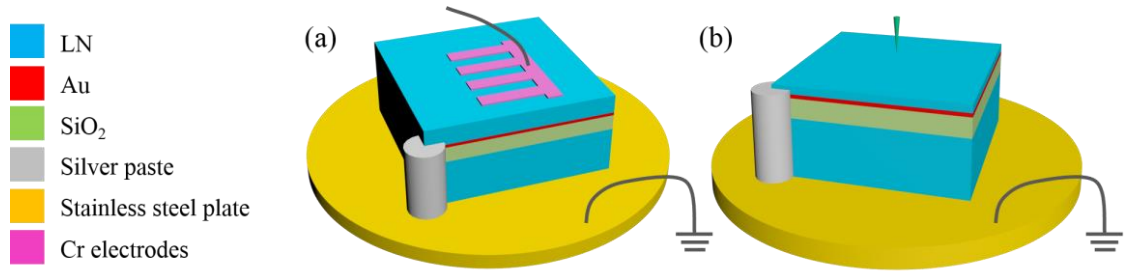

Fig. 1. Schematic diagrams of LNOIs and their corresponding poling techniques for 28-μm (a) and 540-nm (b) thick samples. Blue, red, light green, gray, orange, and pink stand for LN, Pt/Au, SiO$_2$, silver paste, stainless steel plate, and Cr electrodes, respectively. The stainless steel plate is connected to the ground. Positive DC poling voltages are applied to the Cr electrodes in (a) or a conductive tip in (b).

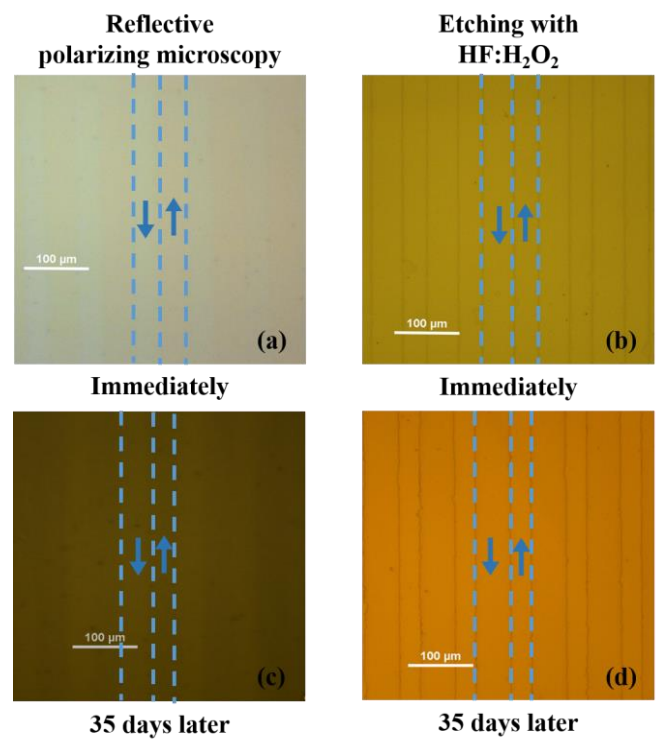

Fig. 2. Domain poling results for 28-μm-thick, 80-μm period PPLNOI. Taken immediately under reflective polarizing microscopy (a), etching with HF:H$_2$O$_2$ (b). Taken 35 days later under reflective polarizing microscopy (c), etching with HF:H$_2$O$_2$ (d), respectively. Up and down arrows indicate positive and negative domains, respectively. Dash lines stand for domain walls.

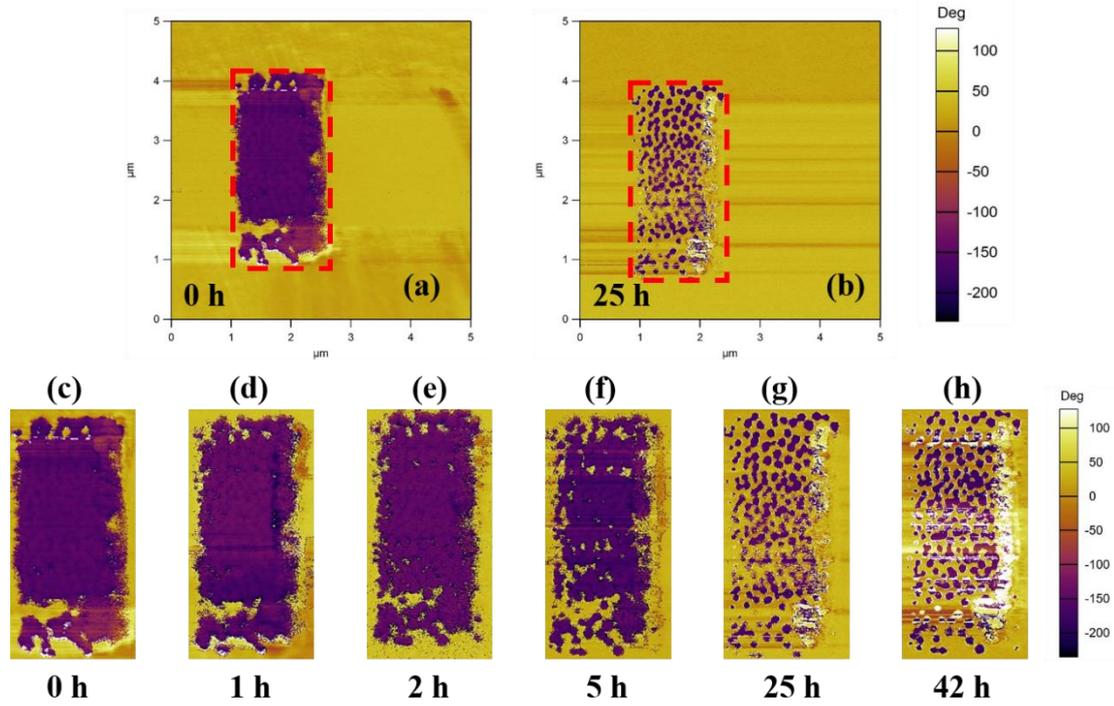

Fig. 3. PFM phase images of the tip-patterned domain. The time intervals between conductive tip scanning and reading are (a) 0 h (immediately) and (b) 25 h. Red rectangles indicate the area over which the positively biased tip scanned. Snapshots of negative domain evolution taken at (c) 0 h, (d) 1 h, (e) 2 h, (f) 5 h, (g) 25 h, and (h) 42 h after poling, respectively.

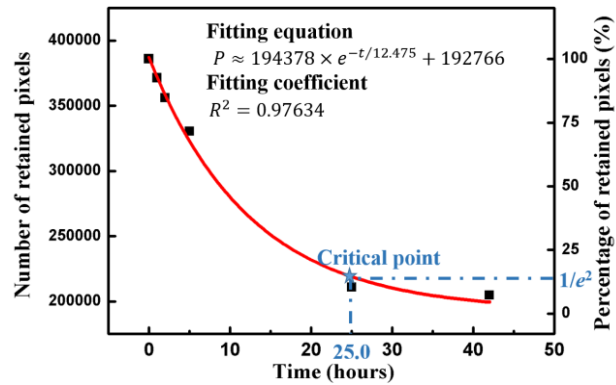

Fig. 4. Retained negative domain area, represented by the number of pixels in the PFM phase images, as a function of time after poling.

8